# Validation of the ultrastable low-noise current amplifier as travelling standard for small direct currents

D. Drung[1], C. Krause[2], S. P. Giblin[3], S. Djordjevic[4], F. Piquemal[4], O. Séron[4], F. Rengnez[4], M. Götz[2], E. Pesel[2], and H. Scherer[2]

[1]Physikalisch-Technische Bundesanstalt (PTB), Abbestraße 2-12, 10587 Berlin, Germany

[2]Physikalisch-Technische Bundesanstalt (PTB), Bundesallee 100, 38116 Braunschweig, Germany

[3]National Physical Laboratory (NPL), Hampton Road, Teddington, Middlesex TW11 0LW, United Kingdom

[4]Laboratoire national de métrologie et d'essais (LNE), 29 avenue Roger Hennequin, 78197 Trappes, France

**Abstract**

An interlaboratory comparison of small-current generation and measurement capability is presented with the ultrastable low-noise current amplifier (ULCA) acting as travelling standard. Various measurements at direct currents between 0.16 nA and 13 nA were performed to verify the degree of agreement between the three national metrology institutes involved in the study. Consistency well within one part per million (ppm) was found. Due to harsh environmental conditions during shipment, the ULCA's transfer accuracy had been limited to about ±0.4 ppm. Supplemental measurements performed at PTB indicate that further improvements in accuracy are possible. Relative uncertainties of 0.1 ppm are achieved by applying on-site calibration of the ULCA with a suitable cryogenic current comparator.

**1. Introduction**

The ultrastable low-noise current amplifier (ULCA) was recently presented as an improved device for the measurement or generation of small direct currents as well as for the calibration of high-value resistors [1, 2]. Its development was stimulated by the need to characterize single-electron transport (SET) devices [3] with 0.1 parts per million (ppm) uncertainty at direct currents of about 100 pA. The use of a cryogenic current comparator (CCC) with a large number of turns as an accurate current amplifier [4] was not pursued at PTB after concerns raised about systematic errors in the CCC at low currents and unavoidable low-frequency excess noise, as explained in detail in [2]. The ULCA concept circumvents such problems: the semiconductor amplifier with excellent low-frequency noise performance is utilized for the measurement of the *small* signal current, but is calibrated with the CCC at *high* currents where the CCC flux linkage is large and rectification due to noise pickup is not a significant concern. The sophisticated ULCA design enables sufficient linearity and stability versus time and temperature.

It was discussed in [2] that using the ULCA for calibrations with currents in the pA range has the potential to lower the uncertainty by up to two orders of magnitude compared to currently established methods. However, all measurements demonstrating the ULCA's superior accuracy so far were performed at PTB with devices operated under stable laboratory conditions and on-site CCC calibration. In this paper, it is shown for the first time how the ULCA's performance is affected by shipment to other laboratories and what level of accuracy is achievable with the ULCA used as a travelling standard. A two-channel unit was carefully studied over a period of about 16 weeks, including transportations to NPL and LNE. Directly before and after each transportation, the ULCA was calibrated at PTB with the standard CCC method [1, 2]. At NPL and LNE, independent calibrations were performed with alternative methods adapted to the existing capabilities. In section 2 the different calibration methods are described in detail. Section 3 presents preparatory measurements done at PTB before the first ULCA transportation. The results of the interlaboratory comparison are reported in Section 4, followed by the conclusions in section 5.

**2. Calibration methods**

The ULCA consists of two stages [1]. The input stage performs current amplification with a matched resistor pair of 3 GΩ and 3 MΩ, built from about 3000 individual chip resistors of 2 MΩ. The nominal current gain is $G_I$ = 1000. The output stage converts the amplified current into a voltage via a 1 MΩ metal-foil resistor network, yielding a nominal output transresistance $R_{IV}$ = 1 MΩ. The total transresistance (i.e., output voltage divided by input current) is nominally $A_{TR}$ = $G_I R_{IV}$ = 1 GΩ. Thus the ULCA basically acts as a current-to-voltage converter.

The standard calibration is performed with PTB's 14-bit CCC [5] in two steps according to the setups depicted in figure 11 of [1]. In the first step, the deviation of the current gain from nominal value $\Delta G_I$ is determined with the CCC at a turns ratio of 16000:16 and currents of ±13 nA and ±13 µA, respectively. The currents are reversed every 10 s, and the averaging time is typically one hour. In the second step, the transresistance of the output stage is calibrated against a 12.9 kΩ standard resistor with a turns ratio of 4029:52 and currents of ±500 nA and ±38.74 µA, respectively, to obtain the deviation from nominal value $\Delta R_{IV}$. The relative standard uncertainties for the calibration of the input and output stages are 0.06 ppm and 0.01 ppm, respectively, including type B contributions [2]. The deviation of the total transresistance from nominal value $\Delta A_{TR}$ and the corresponding overall uncertainty are calculated from the results of the two calibration steps.

As a consistency check, we repeated the two-step calibration using a second CCC, a 12-bit type in place of the 14-bit [6]. For the input stage calibration, the turns ratio had to be lowered to 4000:4 which increased the CCC's current noise level and the required averaging times correspondingly. The total transresistance obtained with the 12-bit CCC was 0.034 ppm smaller than for the 14-bit CCC. However, the type A standard uncertainty of the difference (calculated from the uncertainties of all calibration steps) also was 0.034 ppm, i.e., the observed change was equal to the uncertainty, but well below the quoted calibration uncertainties of 0.061 ppm and 0.091 ppm for the two types of CCCs [2].

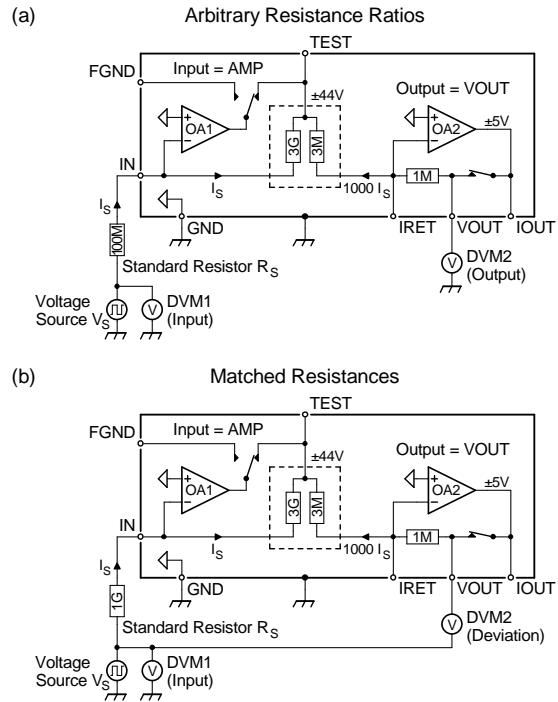

Figure 1. Calibration of the ULCA via a resistance standard $R_S$ (or vice versa, calibration of $R_S$ with the ULCA). Panel (a) shows the general setup for arbitrary resistance ratios $A_{TR}/R_S$, panel (b) a variant for matched resistances $A_{TR} \approx R_S$ that has relaxed requirements on the voltmeter accuracy. Both schemes were applied at NPL with the ULCA set into voltage output mode, i.e., with the internal 1 MΩ resistor performing the current-to-voltage conversion. Alternatively, current output mode may be chosen where the switch between IOUT and VOUT is opened and an external standard resistor is connected between IOUT and IRET [1].

For the ULCA calibrations at NPL, the existing setup for the current measurement of SET devices was adapted [7]. Two temperature-stabilized high-value standard resistors of NPL, 1 GΩ and 100 MΩ (Guildline 9336-series), were calibrated with NPL's 100 V CCC [8]. The quoted standard uncertainties (including type B contributions) are normally 0.8 ppm and 0.2 ppm, respectively. However, the 100 MΩ used in this study was found to have poor short-term stability on time-scales of hours, and an additional uncertainty term was included to allow for drift in between calibrations of the resistor. Following extensive characterization of the resistor, the extra uncertainty was evaluated as a rectangular distribution 0.5 ppm wide, resulting in a total standard uncertainty of 0.25 ppm for the calibration of the 100 MΩ resistor. Additional investigations of two other commercially available 100 MΩ standards suggest that the short-term instability is a generic property of standards based on thick-film resistive elements. Figure 1 depicts two calibration setups comparing the ULCA transresistance $A_{TR}$ against a standard resistor $R_S$. In panel (a), the general scheme for arbitrary resistance ratios $A_{TR}/R_S$ is shown. It was used for the measurements with NPL's 100 MΩ resistor. A voltage $V_S$ is applied to $R_S$ and the resulting current $I_S = V_S/R_S$ is measured with the ULCA. Assuming zero potential at the ULCA input, the resistance ratio is exactly equal to the ratio of the voltmeter readings: $A_{TR}/R_S = V_{DVM2}/V_{DVM1}$.

The setup in figure 1(b) provides improved accuracy for resistance ratios $A_{TR}/R_S \approx 1$, i.e., for $R_S = 1$ GΩ when selecting the voltage output mode. Instead of measuring the ULCA output against ground with DVM2, the deviation between VOUT and the voltage source is metered. This signal is small (zero for $A_{TR} = R_S$) and the demands on the voltmeter accuracy are correspondingly reduced. The resistance ratio is calculated from the voltmeter readings by $A_{TR}/R_S = 1 + V_{DVM2}/V_{DVM1}$. It is worth noting that applying the setup in figure 1(b) for ULCA calibration increases the relative type A uncertainty compared to the two-step calibration with the 14-bit CCC because the input current is lower (±5 nA limited by the ULCA's ±5 V output voltage range instead of ±13 nA with CCC). Furthermore, at low value of $R_S$ Nyquist noise dominates the type A uncertainty (12.8 fA/√Hz for $R_S = 100$ MΩ at 23 °C).

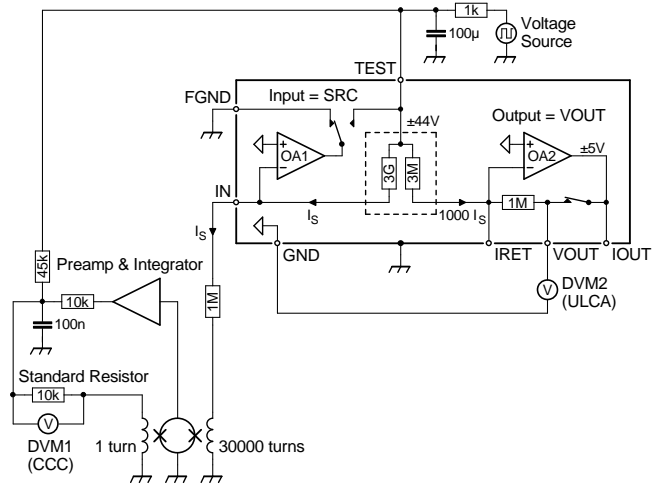

Figure 2. Calibration of the ULCA via a CCC-based current amplifier at LNE. The SQUID is drawn as a circle with two crosses indicating the Josephson junctions. The ULCA is configured as current source [1]. *RC* filters at the outputs of voltage source and integrator are used to limit the slew rate and to reduce noise. DVM1 was operated with an isolation transformer. Its input high side was connected to the CCC winding. To lower the demands on the open-loop gain of the integrator, a 45 kΩ resistor generates a current through the single-turn CCC winding that nominally cancels the effect of the input current $I_S$.

The setup used at LNE is depicted in figure 2. Here, the ULCA is configured as a current source [1], and its output current $I_S$ is measured with LNE's 30000-turn CCC [9]. The flux generated by $I_S$ in the superconducting quantum interference device (SQUID) is cancelled by a servo loop consisting of a preamplifier and integrator (Magnicon XXF-1 SQUID electronics with external integrator option). The feedback current is passed through a 10 kΩ standard resistor into a single-turn CCC coil, thus providing 30000-fold amplification of the current $I_S$. The resulting transresistance of the CCC-based amplifier (i.e., voltage across the 10 kΩ resistor divided by input current $I_S$) amounts to $A_{CCC} = 30000 \times 10$ kΩ = 300 MΩ. The ULCA's transresistance can be calculated from the voltmeter readings: $A_{TR}/A_{CCC} = V_{DVM2}/V_{DVM1}$. Note that DVM2 is connected to the internal ULCA reference potential (triangles in figure 2) instead of the main ground (ULCA housing). Together with a servo loop involving amplifier OA1 this suppresses the effect of the burden voltage between the IN connector and ground [1].

In the ULCA, a 1 pF capacitor is connected in parallel to the 3 GΩ resistor (not shown in figure 2 for clarity) to ensure stability of the input stage amplifier OA1. This results in a high-pass behavior for the noise of the voltage source, and would lead to unacceptably high wideband noise coupled into the SQUID via the 30000-turn winding of the CCC. To avoid this and to limit the slew rate during current reversal, an *RC* filter with a cut-off frequency of about 1.6 Hz was inserted at the output of the voltage source. A further improvement in SQUID stability during current reversal was achieved by adding a 1 MΩ resistor between the ULCA's IN connector and the 30000-turn winding. To lower the demands on the open-loop gain of the integrator, a compensation signal was generated via a resistive divider (45 kΩ and 10 kΩ in figure 2) that nominally cancels the flux in the SQUID caused by the ULCA's output current $I_S$. A 100 nF capacitor against ground provides extra low-pass filtering.

In figure 2 the voltmeter DVM1 is connected in parallel to the 10 kΩ resistor. Therefore, the voltmeter's input resistance could affect the result. To avoid this, instead of measuring the voltage drop across the resistor, two voltmeters could be used to measure the voltage between each resistor terminal and ground. This would remove the effect of the input resistance of the voltmeter at the high-potential side (left resistor terminal in figure 2) by design, and would strongly suppress that of the voltmeter at the low-potential side because the signal at this terminal is about 4 orders of magnitude smaller than that at the high-potential side. For the same reason, the demands on the accuracy of the voltmeter at the low-potential side would be strongly reduced. For the measurements at LNE, however, the simple setup in figure 2 was used because we found that the input resistance of DVM1 (Agilent 3458A on the 10 V range) was about 1 TΩ corresponding to a negligible error of 0.01 ppm.

### 3.    Preparatory measurements at PTB

The calibration methods described in previous section imply the assumption that the voltages at the inputs of the two ULCA stages (IN and IRET in figure 1) are negligibly small, i.e., that the corresponding input resistances $R_1$ and $R_2$ are practically zero. This requires the operational amplifiers OA1 and OA2 to have extremely high open-loop gains, preferably well above $10^9$. The devices used in the ULCA are complex circuits involving several monolithic operational amplifiers and a high-voltage output stage with discrete components, yielding calculated gains of about $3\times10^9$.

To ensure that in practice the gains are sufficiently high, the input resistances $R_1$ and $R_2$ of the two ULCA stages were measured using the PTB CCC bridge electronics and software (without SQUID and CCC). A current of ±10 nA was passed into the ULCA input by one of the two sources, and the voltage between each input terminal (IN or IRET) was measured against ground with the bridge voltage detector. For the latter we selected the chopper amplifier described in [10] because it has negligible voltage noise of about 0.7 nV/√Hz. The input resistance of each stage was calculated by dividing the measured peak-peak voltage by the corresponding peak-peak current (20 nA or 20 µA). The first 2 s after each current reversal were disregarded for the calculation to suppress settling effects. Figure 3 shows the results of an 8¾ hours long input stage measurement and a nearly 6 hours long output stage measurement, $R_1$ = (-0.08 ± 0.24) Ω and $R_2$ = (-0.06 ± 0.03) mΩ, respectively. The input stage's sub-ohm resistance $R_1$ can be considered as a short for the 100 MΩ and 1 GΩ resistors connected to the ULCA input in figure 1. For the output stage, $|R_2| \approx 0.1$ mΩ is negligible compared to the internal 1 MΩ standard resistor, i.e., it is justified that the output voltage in figure 1(a) can be measured with DVM2 against ground instead of across the 1 MΩ resistor.

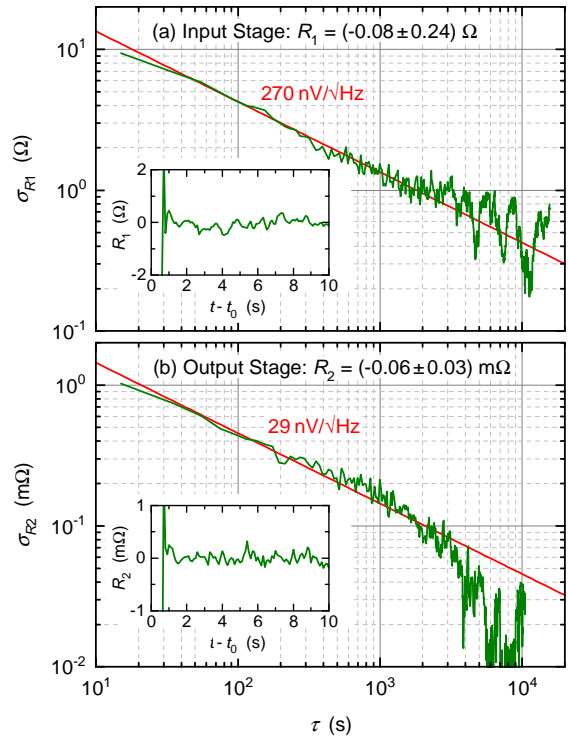

Figure 3. Input resistance of (a) input stage with amplifier OA1 and (b) output stage with amplifier OA2. Main panels show Allan deviations of the voltages at the IN or IRET connectors expressed as resistance, insets show the corresponding settling after current reversal. The white noise levels corresponding to the straight line fits were determined with equation (2) in [1]. They are consistent with the expected voltage noise levels of OA1 and OA2.

The insets in figure 3 demonstrate that the settling of the input voltages is sufficiently fast. Within less than 1 s after current reversal the transients disappear in the noise floor, i.e., voltage spikes at the amplifier inputs during current reversal decay sufficiently fast.

As another important test we compared the CCC calibration in two steps with the calibration via a 100 MΩ standard resistor (MI 9331G). A special ULCA prototype was utilized that is equipped with a 100 kΩ metal-foil resistor at the output stage instead of the usual 1 MΩ. This results in a total transresistance $A_{TR}$ = 100 MΩ, i.e., the less demanding circuit in figure 1(b) can be applied. The setup was slightly modified to enable measurement with the PTB CCC bridge electronics and software (again without SQUID and CCC): The input current $I_S$ was generated by one of the current sources instead of a voltage source, and the deviation signal (DVM2) was measured with the bridge voltage detector. Instead of the chopper amplifier we selected the so-called "bypass" amplifier that has a field-effect transistor input and is preferably used for resistance calibrations above 1 MΩ.

A total of six standard calibrations were performed over a period of three days (cf. figure 4). In between the two standard calibrations of each day, alternative calibrations via the 100 MΩ resistor were done. For this purpose, except at the first day, the 100 MΩ resistor was first calibrated against a 12.9 kΩ standard resistor with the 14-bit CCC at a turns ratio of 15496:2 and currents of ±12.9 nA and ±100 μA, respectively. Subsequently, the ULCA was calibrated with the 100 MΩ resistor at $I_S$ = ±13 nA according to the scheme in figure 1(b). Finally, the CCC calibration of the 100 MΩ resistor was repeated, and the results of the two CCC calibrations were averaged. At the first day, the 100 MΩ measurements were taken in reversed order (ULCA calibrations via 100 MΩ were done before and after a 100 MΩ calibration with CCC). In any case, the selected "time-symmetric" sequence suppresses the effect of slow resistance fluctuations and drift.

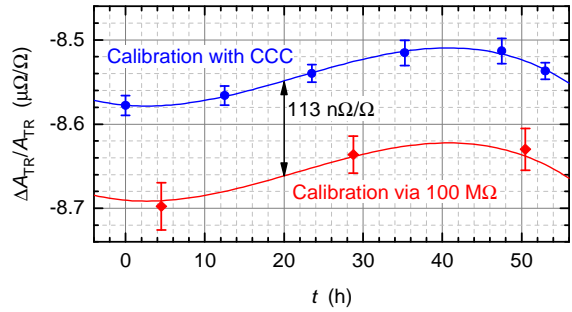

Figure 4. Comparison between the standard CCC calibration in two steps (blue circles) and a calibration via a 100 MΩ resistor according to the setup in figure 1(b) (red diamonds). The measurement was done at PTB with an ULCA having a 100 kΩ output resistor instead of the usual 1 MΩ. The ULCA and the 100 MΩ resistor were placed in a temperature-stabilized air bath. Error bars show type A standard uncertainties. Third-order polynomials (red and blue lines) serve as guides for the eye.

The ULCA's transresistance determined from six CCC calibrations over the 3 day period can be fitted with a third-order polynomial (blue line in figure 4). This polynomial, vertically shifted by 113 nΩ/Ω, matches the ULCA calibrations via the 100 MΩ resistor (red line in figure 4). Thus, we conclude that the two calibration methods are consistent within about 0.1 ppm if the ULCA is operated under laboratory conditions, i.e., without disturbing effects caused by transportation. The slightly smaller calibration value $\Delta A_{TR}/A_{TR}$ obtained with the 100 MΩ resistor could be caused by an extra $9\times10^{14}$ Ω leakage resistance in parallel to the 100 MΩ resistor during its calibration with the 14-bit CCC (e.g., due to leakage in the CCC electronics or wiring). Note that the CCC electronics was originally developed for resistance calibrations of up to 1 MΩ only. The observed effect of 113 nΩ/Ω at 100 MΩ would imply a negligibly small contribution of about 1 nΩ/Ω at 1 MΩ.

## 4. Results of the interlaboratory comparison

Figure 5 summarizes the basic findings of the interlaboratory comparison at PTB, NPL and LNE. For the standard calibrations at PTB, the relative deviations of the input current gain $\Delta G_I/G_I$ and output transresistance $\Delta R_{IV}/R_{IV}$ are depicted in panels (a) and (b), respectively. The relative deviation of the total transresistance $\Delta A_{TR}/A_{TR}$ is plotted in panel (c) along with the NPL and LNE results. To determine the drift of the ULCA over the 16 weeks period, linear fits were calculated for the PTB calibrations (straight lines in figure 5). The drifts in $\Delta G_I/G_I$ and $\Delta R_{IV}/R_{IV}$ ranged between about -1.1 ppm/yr and +0.8 ppm/yr. The total transresistance of ULCA channel A showed an exceptionally low drift of -0.04 ppm/yr because the effects of the input and output stages nearly cancelled. In contrast, for channel B a more typical drift of -2.2 ppm/yr was observed. Temperature effects were corrected for by using the built-in temperature sensors of the two ULCA channels [1]. The displayed ULCA temperature during the measurements at NPL and LNE was about 21.5 °C and 23.8 °C, respectively. Calibrations between 20 °C and 23 °C were performed at PTB, showing no noticeable temperature dependence of the corrected results. During an individual measurement, the ULCA temperature was typically stable within ±0.1 °C.

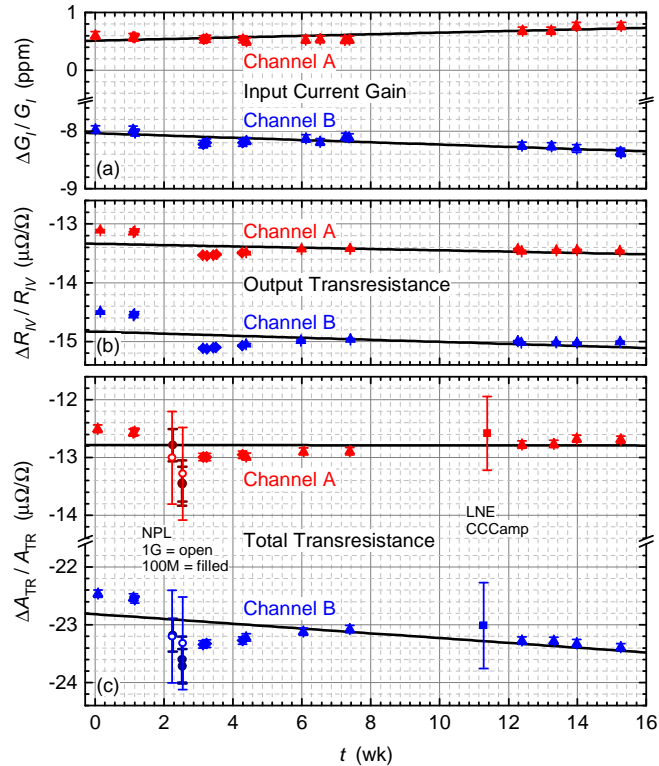

Figure 5. Calibrations of the ULCA at PTB (input stage at ±13 nA, output stage at ±500 nA) along with the measurements at NPL (calibration via 100 MΩ or 1 GΩ at ±5 nA) and LNE (calibration via a CCC-based current amplifier at about ±3 nA). The measurements at NPL and LNE were performed on 11-13 Feb 2015 and 15-17 April 2015, respectively. Directly before and after each shipment, the ULCA was calibrated at PTB. All values are corrected to a nominal temperature of 23°C by applying the known temperature dependencies of the ULCA channels. Error bars show total standard uncertainties including type B contributions ($k = 1$). Solid lines are linear fits through the PTB calibrations.

Figure 5 reveals that the shipment between PTB and NPL, by commercial courier, had a distinct effect on both ULCA channels (see deviations from the linear fits at week 1 and 3). The observed changes in the total transresistances were mainly due to the output stages, and the effect was more pronounced in channel B. Presumably large temperature excursions during air transport have caused the problem, but an influence of mechanical shocks or air pressure changes cannot be excluded. The temperature inside the parcel during the shipment back to PTB ranged from the laboratory temperature of 23 °C down to about 8 °C, which is far below the normal operation temperature. As a precaution, the transfer between PTB and LNE was done by car and monitored by a data logger. During this transfer, the temperature ranged between 17.7 °C and 26 °C, and two vertical shocks of about 160 m/s$^2$ occurred. The "cautious" transfer to LNE did not cause changes clearly distinguishable from the typical drift effects.

The linear fits of $\Delta A_{TR}/A_{TR}$ in figure 5 were used as a reference for comparing the results. The PTB calibrations showed deviations of up to ±0.4 μΩ/Ω shortly before and after the measurements at NPL due to the changes during shipment. After three weeks (i.e., between week 6 and 16 in figure 5) the deviations remained within about ±0.1 μΩ/Ω for both channels. The NPL results (open and filled circles in figure 5) were between zero and 0.8 μΩ/Ω below the linear fits, which is within the standard uncertainty of the resistor calibrations in the case of 1 GΩ, but exceeds the expanded uncertainty (95% confidence, $k = 2$) for two of six 100 MΩ measurements. However, consistency within $k = 2$ is found for all NPL data when comparing with the first PTB calibration after the ULCA's return transport (week 3 in figure 5). As previously noted, the uncertainty in the case of the 100 MΩ measurements was limited by the stability of the resistor used to generate the reference current. The LNE results were about 0.25 μΩ/Ω above the linear fits, i.e., well within the combined standard uncertainty of about 0.7 ppm.

In most cases, the averaging times were chosen long enough to keep the type A uncertainty well below the type B evaluation. Therefore, the error bars in figure 5 are typically dominated by type B effects. The by far largest uncertainty term was the resistor calibration in case of the NPL measurements or the voltmeter calibration in case of the LNE measurements, respectively. In summary, good agreement was found in the interlaboratory comparison. The maximum combined standard uncertainty was about 0.8 μΩ/Ω, and the deviations from the linear fit through the PTB calibrations remained within this value.

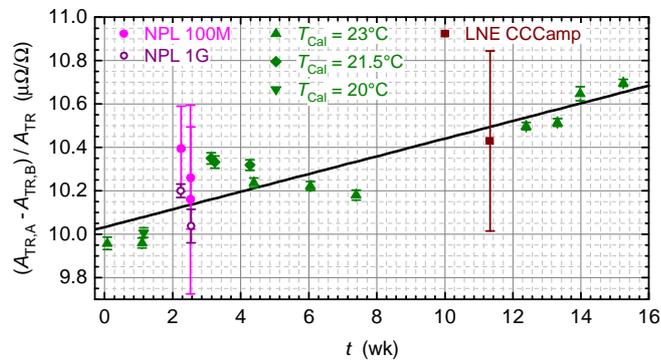

Figure 6. Relative difference in the transresistance of the two ULCA channels taken from figure 5(c). All values are corrected to a nominal temperature of 23°C. The temperature of the PTB calibrations is indicated by different symbols. Error bars show type A standard uncertainties. The solid line is a linear fit through the PTB calibrations. Note that the vertical axis is zoomed compared to figure 5.

To demonstrate the consistency of the data more clearly, the *difference* between the results of the two ULCA channels is plotted in figure 6. The type A uncertainty ("root sum of squares" of the contributions of both channels) is considered here because systematic errors are suppressed by calculating the transresistance difference $A_{TR,A} - A_{TR,B}$. Note that the vertical axis is zoomed compared to figure 5, and that statistical fluctuations around the linear drift line should increase by a factor of √2 compared to the individual ULCA channels. Figure 6 shows good consistency of the results. The deviations between the measured data and the linear fit remain within ±0.2 μΩ/Ω except for one run with the 100 MΩ resistor that exhibits a 0.3 µΩ/Ω excursion. Comparing these deviations with the worst case of -0.8 µΩ/Ω in figure 5(c) and being aware that the two ULCA channels were always measured in direct succession, we conclude that the discrepancies with the 100 MΩ measurements were presumably dominated by the limited stability of the resistor used at NPL.

The ULCA's linearity was investigated with the CCC-based amplifier at LNE (see figure 7). A 14 hours long measurement series was performed with six different current amplitudes $I_P$ between 163 pA and 2.61 nA. The current was reversed every 7 s and the resulting change in the output was referred to the peak-peak amplitude $2I_P$. The type A standard uncertainty for each value of $I_P$ was typically 0.66 fA, corresponding to 2 ppm or 0.13 ppm relative uncertainty at the lowest or highest current, respectively. Figure 7 shows that deviations from the linear behavior remain within the standard uncertainty for 4 of 6 data points, but within the expanded uncertainty (95% confidence, $k = 2$) for all data points. This is the first independent verification of the linearity measurements performed at PTB with the 14-bit CCC [2].

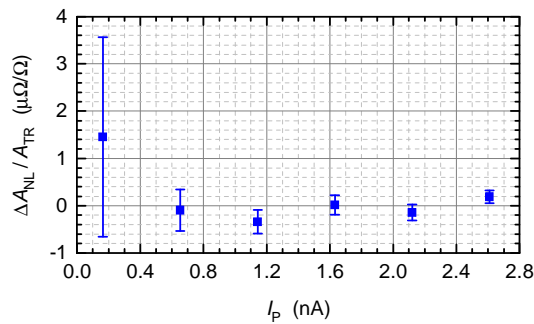

Figure 7. Relative nonlinearity of the transresistance $\Delta A_{NL}/A_{TR}$ of channel A versus peak amplitude $I_P$ measured with LNE's CCC-based current amplifier. $\Delta A_{NL}$ is defined as the difference between the measured transresistance and the arithmetic mean of all data weighted with $I_P$. The amplitude $I_P$ was successively increased from 163 pA to 2.61 nA. Each data point corresponds to a series of 600 full cycles with positive and negative current $\pm I_P$. Error bars show type A standard uncertainties. The nonlinearity of the two voltmeters (Agilent 3458A at 10 V range) is included in the data.

In the setups used for the NPL and LNE comparisons, we selected 10 power line cycles (0.2 s) integration time per voltmeter reading. To remove the voltmeter's input offset, auto-zero was turned on which means that after each reading a zero reading with internally disconnected input signal and shorted amplifier input was taken and then subtracted from the preceding reading. For the LNE measurements, after each current reversal, a 3 s waiting time was inserted to suppress settling effects, and then 10 readings were taken (in total 2 s effective sampling time plus 2 s idle time due to auto-zero). This means that the effective sampling time $\tau_e$ [1] is only 2/7 of the total measurement time. Analyzing the data in figure 7 with equation (2) of Ref. [1] we find an effective current noise level of 23 fA/√Hz at the repetition rate $f_R \approx$ 0.07 Hz. This is slightly larger than the value expected from the noise spectrum of the SQUID in the CCC [9].

The ULCA is preferably read out with an integrating voltmeter in order to avoid aliasing of wideband noise into the measurement bandwidth that could degrade the type A uncertainty. This means that auto-zero should be performed as rarely as possible to minimize the idle time without data acquisition. However, for long intervals between the auto-zero measurements, the voltmeter offset stability can degrade the overall measurement uncertainty. Thanks to the ULCA's high transresistance of 1 GΩ, the requirements on the voltmeter offset stability are relaxed. For the Agilent 3458A voltmeters used in the interlaboratory comparison, subsequent tests showed that it is sufficient to perform auto-zero every 5 s (i.e., after every 24 readings). This reduces the idle time to about 5% of the total measurement time including delays from software triggering and data transmission.

Figure 8 shows the Allan deviations for the two ULCA channels operated with open inputs and read out simultaneously with two voltmeters. A low white noise level of 2.53 fA/√Hz is achieved over a wide range of the sampling time $\tau$ as shown by the dashed line calculated according to [11]. This is only slightly above the intrinsic noise level of 2.4 fA/√Hz [1]. The deviation between fit and experiment at $\tau < 1$ s is due to the increase in the noise spectrum observed above about 1 Hz, probably caused by parasitic capacitance in the 3 GΩ resistor network [1]. At long sampling times the Allan deviation levels off due to $1/f$ noise (the regime where the spectral density $S_I$ scales with $1/f$). From the fit, a $1/f$ corner of about 0.25 mHz is determined (i.e., the frequency at which the white and $1/f$ noise spectral densities are equal). At very long sampling times above about 8 h the Allan deviation of channel A starts to increase, presumably caused by low-frequency excess noise scaling stronger than $1/f$. It was verified by interchanging the two voltmeters that this effect is intrinsic to the ULCA channel.

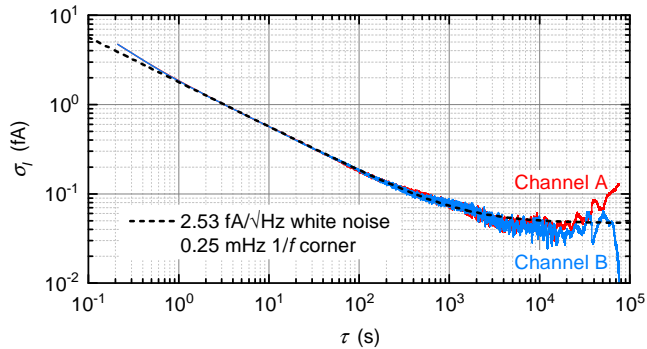

Figure 8. Allan deviation of the input current noise of the two ULCA channels measured simultaneously with two Agilent 3458A voltmeters in 1 V range. An integration time of 10 power line cycles (0.2 s) was selected. Auto-zero was performed after every 24 readings. The time scale was determined from the effective sampling rate of 4.77 readings per second assuming equal intervals between adjacent samples. The dashed line is a fit calculated according to [11] as superposition of white and $1/f$ noise terms.

## 5.    Conclusions

An interlaboratory comparison between PTB, NPL and LNE was successfully performed with a two-channel ULCA acting as travelling standard. Different ULCA measurement schemes were applied to compare the calibration capabilities of the three metrology institutes in the current range around 1 nA. The measurements at NPL and LNE showed standard uncertainties below about 0.8 ppm and agreed well with the PTB calibrations. Deviations within +0.4/-0.8 ppm were found, partially due to changes in the transresistances of the two ULCA channels caused by shipment. Typically, the deviations remained within the standard uncertainty.

Previous interlaboratory comparisons in the small-current regime generally exhibited good agreement between the participants. The accuracy of a recent sub-nanoampere comparison between 13 national metrology institutes was limited by the stability of the picoamperemeters used as travelling instruments [12]. At 100 pA, standard uncertainties of about 10 ppm were obtained at best. Comparisons based on 1 GΩ resistors were performed both at 100 nA [13, 14] and 10 nA [15, 16]. Typically, standard uncertainties of about 5 ppm are achieved with non-cryogenic methods, whereas CCC calibrations allow lower values down to about 1 ppm. However, the available 1 GΩ transfer standards ultimately limit the accuracy of comparisons in the high-resistance (small-current) field.

The use of the ULCA as travelling standard has lead to a substantial improvement, enabling sub-nanoampere interlaboratory comparisons with uncertainties below 1 ppm. It is worth noting that the ULCA's shipment caused noticeable changes in the internal 1 MΩ resistors of the output stages only, whereas the current gains of the input stages remained virtually unaffected – see figure 5(a). Therefore, the application of an external calibrated 1 MΩ standard resistor (as typically available in metrology institutes) could reduce the transfer uncertainty of the ULCA to nearly 0.1 ppm, i.e., almost the same accuracy could be achieved that is presently obtained if the ULCA is operated under laboratory conditions and calibrated on-site with the 14 bit CCC shortly before and after a current measurement.

The presented findings are highly relevant for high-accuracy measurements on SET current sources. Recently, at PTB, current quantization on SET current sources at pump frequencies up to 1 GHz range was validated with uncertainties down to 0.2 ppm by using the ULCA under laboratory conditions with on-site CCC calibration [17]. Achieving nearly the same uncertainty with the ULCA as travelling standard as discussed above would thus enable interlaboratory comparisons of SET pumps at highest accuracy, valuable for universality tests towards the implementation of true SET-based quantum current standards.

**Acknowledgements**

The authors would like to thank a number of people for their valuable contribution to the interlaboratory comparison: Colin Porter, Ben Thornton and Clinton Kelly at NPL for assistance with resistance and voltage calibrations, Olivier Thévenot at LNE for loaning the 10 kΩ resistance standard, Alain Pesel and Wilfrid Poirier at LNE for resistance calibrations, Michael Piepenhagen at PTB for fabrication of printed-circuit boards, and Friederike Stein for stimulating discussions. This work was supported by the Joint Research Project "Qu-Ampere" (SIB07) within the European Metrology Research Programme (EMRP). The EMRP is jointly funded by the EMRP participating countries within EURAMET and the European Union.